\documentclass{PoS}

\title{The dark matter is mostly an axion BEC }

\ShortTitle{The dark matter is mostly an axion BEC }

\author{\speaker{Pierre SIKIVIE}%
         \thanks{I thank Ozgur Erken, Heywood Tam and Qiaoli Yang 
for useful discussions.  This work was supported in part by the 
US Department of Energy under contract DE-FG02-97ER41029.}\\
        University of Florida\\
        E-mail: \email{sikivie@phys.ufl.edu}}

\abstract{Axions differ from ordinary cold dark matter, such as 
WIMPs or sterile neutrinos, because they form a Bose-Einstein 
condensate (BEC).  As a result, axions accreting onto a galactic 
halo fall in with net overall rotation.  In contrast, ordinary CDM 
accretes onto galactic halos with an irrotational velocity field.  
The inner caustics are different in the two cases.  It is shown that 
if the dark matter is axions, the phase space structure of the halos 
of isolated disk galaxies, such as the Milky Way, is precisely that 
of the caustic ring model for which observational support exists.  
The other dark matter candidates predict a far more chaotic phase 
space structure for galactic halos.}

\FullConference{Identification of Dark Matter 2010\\
                 July 26 - 30 2010\\
                 University of Montpellier 2, Montpellier, France}

\begin{document}

\section{Introduction}

One of the outstanding problems in science today is the identity
of the dark matter of the universe \cite{PDM}.  The existence of   
dark matter is implied by a large number of observations, including
the dynamics of galaxy clusters, the rotation curves of individual
galaxies, the abundances of light elements, gravitational lensing,
and the anisotropies of the cosmic microwave background radiation.
The energy density fraction of the universe in dark matter is 23\%.
The dark matter must be non-baryonic, cold and collisionless.  {\it Cold}
means that the primordial velocity dispersion of the dark matter particles
is sufficiently small, less than about $10^{-8}~c$ today, so that it may 
be set equal to zero as far as the formation of large scale structure and
galactic halos is concerned.  {\it Collisionless} means that the dark matter
particles have, in first approximation, only gravitational interactions.    
Particles with the required properties are referred to as `cold dark matter'
(CDM).  The leading CDM candidates are weakly interacting massive particles 
(WIMPs) with mass in the 100 GeV range, axions with mass in the $10^{-5}$ eV
range, and sterile neutrinos with mass in the keV range.  

Today I argue that the dark matter is axions \cite{CABEC,case}.  The 
argument has three parts.  First, axions behave differently from the 
other forms of cold dark matter because they form a Bose-Einstein 
condensate \cite{CABEC}.  Second, there is a tool to distinguish axion 
BEC from the other forms of CDM on the basis of observation, namely the 
study of the inner caustics of galactic halos.  Third, the evidence for 
caustic rings of dark matter is consistent in every aspect with axion BEC, 
but not with WIMPs or sterile neutrinos.

Before I start, let me mention that H. Baer and his collaborators have shown
that in many supersymmetric extensions of the Standard Model, the dark matter 
is axions, entirely or in part \cite{Baer}. 

\section{Axions}

Shortly after the Standard Model of elementary particles was 
established, the axion was postulated \cite{axion} to explain 
why the strong interactions conserve the discrete symmetries P 
and CP.  For our purposes the action density for the axion field
$\varphi(x)$ may be taken to be
\begin{equation}
{\cal L}_a = - {1 \over 2} \partial_\mu \varphi \partial^\mu \varphi
- {1 \over 2} m^2 \varphi^2 + {\lambda \over 4!} \varphi^4 - ...
\label{lag}
\end{equation}
where $m$ is the axion mass.  The self-coupling strength is
\begin{equation}
\lambda = {m^2 \over f^2}~{m_d^3 + m_u^3 \over (m_d + m_u)^3}
\simeq 0.35~{m^2 \over f^2}
\label{self}
\end{equation}
in terms of the axion decay constant $f$ and the masses $m_u$ and
$m_d$ of the up and down quarks.  In Eq.~(\ref{lag}), the dots
represent higher order axion self-interactions and interactions
of the axion with other particles.  All axion couplings and the
axion mass
\begin{equation}
m \simeq 6 \cdot 10^{-6}~{\rm eV}~{10^{12}~{\rm GeV} \over f}
\label{mass}
\end{equation}
are inversely proportional to $f$. $f$ was first thought to be of order
the electroweak scale, but its value is in fact arbitrary \cite{invis}.  
However, the combined limits from unsuccessful searches in particle
and nuclear physics experiments and from stellar evolution require
$f \gtrsim 3 \cdot 10^9$ GeV \cite{axrev}.

Furthermore, an upper limit $f \lesssim 10^{12}$ GeV is provided by
cosmology because light axions are abundantly produced during the QCD
phase transition \cite{axdm}.  In spite of their very small mass, these
axions are a form of cold dark matter.  Indeed, their average momentum
at the QCD epoch is not of order the temperature (GeV) but of order the
Hubble expansion rate ($3 \cdot 10^{-9}$ eV) then.  In case inflation
occurs after the Peccei-Quinn phase transition their average momentum is
even smaller because the axion field gets homogenized during inflation.
For a detailed discussion see ref. \cite{axcos}.  In addition to this
cold axion population, there is a thermal axion population with average
momentum of order the temperature.

The non-perturbative QCD effects that give the axion its mass turn on
at a temperature of order 1 GeV.  The critical time, defined by
$m(t_1) t_1 = 1$, is $t_1 \simeq 2 \cdot 10^{-7}~{\rm sec}~
(f / 10^{12}~{\rm GeV})^{1 \over 3}$.  Cold axions are the quanta
of oscillation of the axion field that result from the turn on of
the axion mass.  They have number density
\begin{equation}
n(t) \sim {4 \cdot 10^{47} \over {\rm cm}^3}~
\left({f \over 10^{12}~{\rm GeV}}\right)^{5 \over 3}
\left({a(t_1) \over a(t)}\right)^3
\label{numden}
\end{equation}
where $a(t)$ is the cosmological scale factor.  Because the axion 
momenta are of order ${1 \over t_1}$ at time $t_1$ and vary with
time as $a(t)^{-1}$, the velocity dispersion of cold axions is
\begin{equation}
\delta v (t) \sim {1 \over m t_1}~{a(t_1) \over a(t)}
\label{veldis}
\end{equation}
{\it if} each axion remains in whatever state it is in, i.e. if axion
interactions are negligible.  Let us refer to this case as the limit of
decoupled cold axions.  If decoupled, the average state occupation number 
of cold axions is
\begin{equation}
{\cal N} \sim~ n~{(2 \pi)^3 \over {4 \pi \over 3} (m \delta v)^3}
\sim 10^{61}~\left({f \over 10^{12}~{\rm GeV}}\right)^{8 \over 3}~~\ .
\label{occnum}  
\end{equation}
Clearly, the effective temperature of cold axions is much smaller than
the critical temperature
\begin{equation}
T_{\rm c} = \left({\pi^2 n \over \zeta(3)}\right)^{1 \over 3}
\simeq 300~{\rm GeV}~\left({f \over 10^{12}~{\rm GeV}}\right)^{5 \over 9}~
{a(t_1) \over a(t)}
\label{Tc}
\end{equation}
for Bose-Einstein condensation.  Bose-Einstein (BEC) may be briefly 
described as follows: if identical bosonic particles are highly condensed 
in phase space, if their total number is conserved and if they thermalize, 
most of them go to the lowest energy available state.  The condensing 
particles do so because, by yielding their energy to the remaining 
non-condensed particles, the total entropy is increased.  

Eqs.~(\ref{occnum}) and (\ref{Tc}) tell us that the first condition 
is overwhelmingly satisfied.  The second condition is also satisfied 
because all axion number violating processes, such as their decay to 
two photons, occur on time scales vastly longer than the age of the 
universe.  The only condition for axion BEC that is not manifestly 
satisfied is thermal equilibrium.  Thermal equilibrium of axions may 
seem unlikely because the axion is very weakly coupled.  However, it 
was found in ref. \cite{CABEC} that dark matter axions do form a BEC, 
marginally because of their self-interactions, but certainly as a result 
of their gravitational interactions.  No special assumptions are required.

\section{Bose-Einstein condensation of cold dark matter axions}

Axions are in thermal equilibrium if their relaxation rate $\Gamma$ is
large compared to the Hubble expansion rate $H(t) = {1 \over 2t}$.  At  
low phase space densities, the relaxation rate is of order the particle  
interaction rate $\Gamma_s = n \sigma \delta v$ where $\sigma$ is the
scattering cross-section.  The cross-section for
$\varphi + \varphi \rightarrow \varphi + \varphi$ scattering due to
axion self interaction is {\it in vacuum}
\begin{equation}
\sigma_0 = {1 \over 64 \pi} {\lambda^2 \over m^2} \simeq
1.5 \cdot 10^{-105} {\rm cm}^2 \left({m \over 10^{-5}~{\rm eV}}\right)^6~~~\ .
\label{xs0}
\end{equation}
If one substitutes $\sigma_0$ for $\sigma$, $\Gamma_s$ is found much
smaller than the Hubble rate, by many orders of magnitude.  However,
in the cold axion fluid background, the scattering rate is enhanced by
the average quantum state occupation number of both final state axions,
$\sigma \sim \sigma_0 {\cal N}^2$, because energy conservation forces
the final state axions to be in highly occupied states if the initial
axions are in highly occupied states.  In that case, the relaxation rate
is multiplied by {\it one} factor of ${\cal N}$ \cite{ST}
\begin{equation}
\Gamma \sim n~\sigma_0~\delta v~{\cal N}~~~\ .
\label{rate}
\end{equation}
Combining Eqs.~(\ref{numden}-\ref{occnum},\ref{xs0}), one finds
$\Gamma(t_1)/H(t_1) \sim {\cal O}(1)$, suggesting that cold axions
thermalize at time $t_1$ through their self interactions, but only   
barely so.

A critical aspect of axion BEC phenomenology is whether the BEC
continues to thermalize after it has formed.  Axion BEC means
that (almost) all axions go to one state.  However, only if the
BEC continually rethermalizes does the axion state track the
lowest energy state.

The particle kinetic equations that yield Eq.~(\ref{rate}) are
valid only when the energy dispersion ${1 \over 2} m (\delta v)^2$
is larger than the thermalization rate \cite{ST}.  After $t_1$ this   
condition is no longer satisfied.  One enters then a regime where the 
relaxation rate due to self interactions is of order \cite{CABEC,four}
\begin{equation}
\Gamma_\lambda  \sim \lambda~n~m^{-2}~~\ .
\label{rate2}
\end{equation}
$\Gamma_\lambda(t)/H(t)$ is of order one at time $t_1$ but
decreases as $t~a(t)^{-3}$ afterwards.  Hence, self interactions
are insufficient to cause axion BEC to rethermalize after $t_1$
even if they cause axion BEC at $t_1$.  However gravitational
interactions, which are long range, do the job later on.  The
relaxation rate due to gravitational interactions is of order
\cite{CABEC,four}
\begin{equation}
\Gamma_{\rm g} \sim G~n~m^2~l^2
\label{rate3}
\end{equation}
where $l \sim (m \delta v)^{-1}$ is the correlation length.
$\Gamma_{\rm g}(t)/H(t)$ is of order
$4 \cdot 10^{-8}(f/10^{12}~{\rm GeV})^{2 \over 3}$
at time $t_1$ but grows as $t a^{-1}(t) \propto a(t)$.  Thus
gravitational interactions cause the axions to thermalize and
form a BEC when the photon temperature is of order
100 eV~$(f/10^{12}~{\rm GeV})^{1 \over 2}$.

The process of axion Bose-Einstein condensation is constrained by 
causality.  We expect overlapping condensate patches with typical    
size of order the horizon.  As time goes on, say from $t$ to $2t$,
the axions in $t$-size condensate patches rethermalize into $2t$-size
patches.  The correlation length is then of order the horizon at   
all times, implying $\delta v \sim {1 \over m t}$ instead of
Eq.~(\ref{veldis}), and $\Gamma_{\rm g}/H \propto t^3 a^{-3}(t)$
after the BEC has formed.  Therefore gravitational interactions  
rethermalize the axion BEC on ever shorter time scales compared
to the age of the universe.  The question now is whether axion BEC 
has implications for observation.

\section{Dark matter caustics}

The study of the inner caustics of galactic halos \cite{crdm,inner}
may provide a useful tool.  An isolated galaxy like our own accretes 
the dark matter particles surrounding it.  Cold collisionless particles 
falling in and out of a gravitational potential well necessarily form an 
inner caustic, i.e. a surface of high density, which may be thought of as 
the envelope of the particle trajectories near their closest approach to 
the center.  The density diverges at caustics in the limit where the
velocity dispersion of the dark matter particles vanishes.  Because 
the accreted dark matter falls in and out of the galactic gravitational
potential well many times, there is a set of inner caustics.  In addition,
there is a set of outer caustics, one for each outflow as it reaches its
maximum radius before falling back in.  We will be concerned here with
the catastrophe structure and spatial distribution of the inner caustics
of isolated disk galaxies.

The catastrophe structure of the inner caustics depends mainly on the angular
momentum distribution of the infalling particles \cite{inner}.  There are
two contrasting cases to consider.  In the first case, the angular momentum
distribution is characterized by `net overall rotation';  in the second case,
by irrotational flow.  The archetypical example of net overall rotation is 
instantaneous rigid rotation on the turnaround sphere.  The turnaround sphere
is defined as the locus of particles which have zero radial velocity with
respect to the galactic center for the first time, their outward Hubble flow
having just been arrested by the gravitational pull of the galaxy.  The
present turnaround radius of the Milky Way is of order 2 Mpc.  Net overall 
rotation implies that the velocity field has a curl, $\vec{\nabla} \times  
\vec{v} \neq 0$.  The corresponding inner caustic is a closed tube whose
cross-section is a section of the elliptic umbilic ($D_{-4})$ catastrophe  
\cite{crdm,inner}.  It is often referred to as a `caustic ring', or
`tricusp ring' in reference to its shape.  In the case of irrotational
flow, $\vec{\nabla} \times \vec{v} = 0$, the inner caustic has a tent-like
structure quite distinct from a caustic ring.  Both types of inner caustic
are described in detail in ref.\cite{inner}.

If a galactic halo has net overall rotation and its time evolution
is self-similar, the radii of its caustic rings are predicted in   
terms of a single parameter, called $j_{\rm max}$.  Self-similarity
means that the entire phase space structure of the halo is time
independent except for a rescaling of all distances by $R(t)$, all
velocities by $R(t)/t$ and all densities by $1/t^2$ \cite{FG,B,STW,MWh}. 
For definiteness, $R(t)$ will be taken to be the turnaround radius at 
time $t$.  If the initial overdensity around which the halo forms has 
a power law profile
\begin{equation}
{\delta M_i \over M_i} \propto ({1 \over M_i})^\epsilon~~~\ ,
\label{inov}
\end{equation}
where $M_i$ and $\delta M_i$ are respectively the mass
and excess mass within an initial radius $r_i$, then
$R(t) \propto t^{{2 \over 3} + {2 \over 9 \epsilon}}$ \cite{FG}.
In an average sense, $\epsilon$ is related to the slope of the evolved
power spectrum of density perturbations on galaxy scales \cite{Dor}.
The observed power spectrum implies that $\epsilon$ is in the range
0.25 to 0.35 \cite{STW}.  The prediction for the caustic ring radii
is ($n$ = 1, 2, 3, .. )
\cite{crdm,MWh}
\begin{equation}
a_n \simeq {{\rm 40~kpc} \over n}~
\left({v_{\rm rot} \over 220~{\rm km/s}}\right)~
\left({j_{\rm max} \over 0.18}\right)
\label{crr}
\end{equation}
where $v_{\rm rot}$ is the galactic rotation velocity.
Eq.(~\ref{crr}) is for $\epsilon = 0.3$.  The $a_n$ have a
small $\epsilon$ dependence.  However, the $a_n \propto 1/n$
approximate behavior holds for all $\epsilon$ in the range 0.25
and 0.35, so that a change in $\epsilon$ is equivalent to a change
in $j_{\rm max}$.  $(\epsilon, j_{\rm max})$ = (0.30, 0.180) implies
very nearly the same radii as $(\epsilon, j_{\rm max})$ =  (0.25, 0.185)  
and (0.35, 0.177).

Observational evidence for caustic rings with the radii predicted by
Eq.~(\ref{crr}) was found in the statistical distribution of bumps
in a set of 32 extended and well-measured galactic rotation curves
\cite{Kinn}, the distribution of bumps in the rotation curve of the
Milky Way \cite{milky}, the appearance of a triangular feature in
the IRAS map of the Milky Way in the precise direction tangent to   
the nearest caustic ring \cite{milky}, and the existence of a ring  
of stars at the location of the second ($n$ = 2) caustic ring in   
the Milky Way \cite{Mon}.  Each galaxy may have its own value of  
$j_{\rm max}$.  However, the $j_{\rm max}$ distribution over the
galaxies involved in the aforementioned evidence is found to be
peaked at 0.18.  There is evidence also for a caustic ring of
dark matter in a galaxy cluster \cite{clu}.

Recently the rotation curve of our nearest large neighbour, the 
Andromeda galaxy, was measured with far greater precision and detail 
than hitherto achieved \cite{Chemin}.  The new rotation curve has 
three prominent bumps, at 10 kpc, 15 kpc and 29 kpc.  The positions 
of these bumps are in the ratios predicted by the caustic ring model 
and thus provide fresh additional evidence.
  
\section{The caustic ring halo model}

The caustic ring  model of galactic halos \cite{MWh} is the phase space 
structure that follows from self-similarity, axial symmetry, and net overall 
rotation.  Self-similarity requires that the time-dependence of the specific
angular momentum distribution on the turnaround sphere be given by  
\cite{STW,MWh}
\begin{equation}
\vec{\ell}(\hat{n},t) = \vec{j}(\hat{n})~{R(t)^2 \over t}
\label{td}    
\end{equation}
where $\hat{n}$ is the unit vector pointing to a position
on the turnaround sphere, and $\vec{j}(\hat{n})$ is a
dimensionless time-independent angular momentum distribution.   
In case of instantaneous rigid rotation, which is the simplest
form of net overall rotation,
\begin{equation}
\vec{j}(\hat{n}) = j_{\rm max}~\hat{n} \times (\hat{z} \times \hat{n})
\label{irr}
\end{equation}  
where $\hat{z}$ is the axis of rotation and $j_{\rm max}$ is the
parameter that appears in Eq.~(\ref{crr}).  The angular velocity
is $\vec{\omega} = {j_{\rm max} \over t} \hat{z}$.  Each property
of the assumed angular momentum distribution maps onto an observable
property of the inner caustics: net overall rotation causes the inner
caustics to be rings, the value of $j_{\rm max}$ determines their
overall size, and the time dependence given in Eq.~(\ref{td}) causes
$a_n \propto 1/n$.

The angular momentum distribution assumed by the caustic ring  
halo model may seem implausible because it is highly organized
in both time and space.  Numerical simulations \cite{num} suggest
that galactic halo formation is a far more chaotic process.  However, 
since the model is motivated by observation, it is appropriate to ask
whether it is consistent with the expected behaviour of some or any of 
the dark matter candidates.  In addressing this question we make the 
usual assumption, commonly referred to as `tidal torque theory', that 
the angular momentum of a galaxy is due to the tidal torque applied to 
it by nearby protogalaxies early on when density perturbations are still 
small and protogalaxies close to one another \cite{ttt,Peeb}.  We divide 
the question into three parts:  1. is the value of $j_{\rm max}$ consistent 
with the magnitude of angular momentum expected from tidal torque theory?  
2. is it possible for tidal torque theory to produce net overall rotation? 
3. does the axis of rotation remain fixed in time, and is Eq.~(\ref{td}) 
expected as an outcome of tidal torque theory?

\section{Magnitude of angular momentum}

The amount of angular momentum acquired by a galaxy through tidal 
torquing can be reliably estimated by numerical simulation because
it does not depend on any small feature of the initial mass configuration,
so that the resolution of present simulations is not an issue in this case.
The dimensionless angular momentum parameter
\begin{equation}
\lambda \equiv {L |E|^{1 \over 2} \over G M^{5 \over 2}}~~\ ,
\label{lambda}
\end{equation}
where $G$ is Newton's gravitational constant, $L$ is the angular
momentum of the galaxy, $M$ its mass and $E$ its net mechanical
(kinetic plus gravitational potential) energy, was found to have
median value 0.05 \cite{Efst}.  In the caustic ring model the   
magnitude of angular momentum is given by $j_{\rm max}$.  As  
mentioned, the evidence for caustic rings implies that the
$j_{\rm max}$-distribution is peaked at $j_{\rm max} \simeq$ 0.18.
Is the value of $j_{\rm max}$ implied by the evidence for caustic
rings compatible with the value of $\lambda$ predicted by tidal
torque theory?  

The relationship between $j_{\rm max}$ and $\lambda$ may be easily
derived.  Self-similarity implies that the halo mass $M(t)$ within
the turnaround radius $R(t)$ grows as $t^{2 \over 3 \epsilon}$ \cite{FG}.
Hence the total angular momentum grows according to
\begin{equation}
{d \vec{L} \over dt} = \int d\Omega {dM \over d\Omega dt} \vec{\ell}
= {4 \over 9 \epsilon} {M(t) R(t)^2 \over t^2} j_{\rm max}~\hat{z}
\label{grL}
\end{equation}
where we assumed, for the sake of definiteness, that the infall is
isotropic and that $\vec{j}(\hat{n})$ is given by Eq. (\ref{irr}).
Integrating Eq.~(\ref{grL}), we find
\begin{equation}
\vec{L}(t) = {4 \over 10 + 3 \epsilon}~
{M(t) R(t)^2 \over t} j_{\rm max}~\hat{z}~~~\ .
\label{Lt}
\end{equation}
Similarly, the total mechanical energy is
\begin{equation}
E(t) = - \int {G M(t) \over R(t)} {dM \over dt} dt =
- {3 \over 5 - 3 \epsilon} {G M(t)^2 \over R(t)}~~~\ .
\label{Et}
\end{equation}
Here we use the fact that each particle on the turnaround sphere has  
potential energy $- G M(t)/R(t)$ and approximately zero kinetic energy.
Combining Eqs. (\ref{lambda}), (\ref{Lt}) and (\ref{Et}) and using the
relation $R(t)^3 = {8 \over \pi^2} t^2 G M(t)$
\cite{FG}, we find
\begin{equation}
\lambda = \sqrt{6 \over 5 - 3 \epsilon}~
{8 \over 10 + 3 \epsilon}~
{1 \over \pi}~j_{\rm max}~~~\ .
\label{rel}
\end{equation}  
For $\epsilon$ = 0.25, 0.30 and 0.35, Eq.~(\ref{rel}) implies
$\lambda/j_{\rm max}$ = 0.281, 0.283 and 0.284 respectively. Hence
there is excellent agreement between $j_{\rm max} \simeq 0.18$
and $\lambda \sim 0.05$.

The agreement between $j_{\rm max}$ and $\lambda$ gives further 
credence to the caustic ring model.  Indeed if the evidence for 
caustic rings were incorrectly interpreted, there would be no 
reason for it to produce a value of $j_{\rm max}$ consistent
with $\lambda$.  Note that the agreement is excellent only in
Concordance Cosmology.  In a flat matter dominated universe,
the value of $j_{\rm max}$ implied by the evidence for caustic 
rings is 0.27 \cite{crdm,MWh}.

\section{Net overall rotation}

Next we ask whether net overall rotation is an expected
outcome of tidal torquing.  The answer is clearly no if
the dark matter is collisionless.  Indeed, the velocity
field of collisionless dark matter satisfies
\begin{equation}
{d \vec{v} \over dt}(\vec{r}, t) =
{\partial \vec{v} \over \partial t}(\vec{r}, t) +
(\vec{v}(\vec{r}, t) \cdot \vec{\nabla}) \vec{v} (\vec{r}, t)
= - \vec{\nabla} \phi(\vec{r}, t)
\label{cdm}
\end{equation}  
where $\phi(\vec{r}, t)$ is the gravitational potential.  The
initial velocity field is irrotational because the expansion
of the universe caused all rotational modes to decay away
\cite{denp}.  Furthermore, it is easy to show \cite{inner}
that if $\vec{\nabla} \times \vec{v} = 0$ initially, then
Eq.~(\ref{cdm}) implies $\vec{\nabla} \times \vec{v} = 0$
at all later times.  Since net overall rotation requires
$\vec{\nabla} \times \vec{v} \neq 0$, it is inconsistent with
collisionless dark matter, such as WIMPs or sterile neutrinos.
If WIMPs or sterile neutrinos are the dark matter, the evidence
for caustic rings, including the agreement between $j_{\rm max}$
and $\lambda$ obtained above, is purely fortuitous.

Axions \cite{axion,invis,axdm,axrev} differ from WIMPs and sterile 
neutrinos.  Axions are not collisionless, in the sense of Eq.~(\ref{cdm}), 
because they form a rethermalizing Bose-Einstein condensate.  This process 
is quantum mechanical in an essential way and not described by Eq.~(\ref{cdm}). 
By {\it rethermalizing} we mean that thermalization rate remains larger than 
the Hubble rate so that the axion state tracks the lowest energy available 
state. The compressional (scalar) modes of the axion field are unstable 
and grow as for ordinary CDM, except on length scales too small to be of 
observational interest \cite{CABEC}.  Unlike ordinary CDM, however, the 
rotational (vector) modes of the axion field exchange angular momentum 
by gravitational interaction.  Most axions condense into the state 
of lowest energy consistent with the total angular momentum, say 
$\vec{L} = L \hat{z}$, acquired by tidal torquing at a given time.  
To find this state we may use the WKB approximation because the 
angular momentum quantum numbers are very large, of order $10^{20}$ 
for a typical galaxy.  The WKB approximation maps the axion wavefunction 
onto a flow of classical particles with the same energy and momentum 
densities.  It is easy to show that for given total angular momentum 
the lowest energy is achieved when the angular motion is rigid rotation.  
So we find Eq.~(\ref{irr}) to be a prediction of tidal torque theory 
if the dark matter is axions.

Thermalization by gravitational interactions is only effective between
modes of very low relative momentum because only in this case is the 
correlation length $l$, that appears in Eq.~(\ref{rate3}), large.  After 
the axions fall into the gravitational potential well of the galaxy, they 
form multiple streams and caustics like ordinary CDM \cite{WK}.  The momenta 
of particles in different streams are too different from each other for 
thermalization by gravitational interactions to occur across streams.  
The wavefunction of the axions inside the turnaround sphere is mapped by 
the WKB approximation onto the flow of classical particles with the same 
initial conditions on that sphere.  The phase space structure thus formed 
has caustic rings since the axions reach the turnaround sphere with net 
overall rotation.  The axion wavefunction vanishes on an array of lines. 
These lines, numbering of order $10^{20}$, may be thought of as the vortices 
characteristic of a BEC with angular momentum.  However, the transverse size 
of the axion vortices is of order the inverse momentum associated with the 
radial motion in the halo, $(m v_r)^{-1} \sim$ 20 meters for a typical value 
($10^{-5}$ eV) of the axion mass.  In a BEC without radial motion the size of 
vortices is of order the healing length \cite{PJ}, which is much larger than 
$(m v_r)^{-1}$.

One might ask whether there is a way in which net overall rotation
may be obtained other than by BEC of the dark matter particles.  I
could not find any.  General relativistic effects may produce a curl
in the velocity field but are only of order $(v/c)^2 \sim 10^{-6}$ 
which is far too small for the purposes described here.  One may  
propose that the dark matter particles be collisionfull in the sense
of having a sizable cross-section for elastic scattering off each other.
The particles then share angular momentum by particle collisions after
they have fallen into the galactic gravitational potential well.  However,
the collisions fuzz up the phase space structure that we are trying to
account for.  The angular momentum is only fully shared among the halo
particles after the flows and caustics of the model are fully destroyed.
Axions appear singled out in their ability to produce the net overall
rotation implied by the evidence for caustic rings of dark matter.

If the dark matter is WIMPs or sterile neutrinos, the velocity field
of dark matter is curl-free.  As already mentioned, the inner caustics 
of galactic halos have then a tent-like structure which is quite distinct 
from caustic rings \cite{inner}.  Also, the angular momentum of the dark 
matter accreting onto a halo is not shared among the infalling particles.  
The total angular momentum vector $\vec{L}$ of the halo is the same as 
for axion dark matter, since it is determined by the outcome of tidal 
torque theory, but unlike the axion case is the sum of many contributions 
randomly oriented with respect to one another.   The tent-like inner 
caustics have therefore random orientations, whereas the caustic rings 
of the axion case lie all in the galactic plane.

\section{Self-similarity}

The third question provides a test of the conclusions reached
so far.  If galaxies acquire their angular momentum by tidal
torquing and if the dark matter particles are axions in a
rethermalizing Bose-Einstein condensate, then the time dependence
of the specific angular momentum distribution on the turnaround   
sphere is predicted.   Is it consistent with Eq.~(\ref{td})?
In particular, is the axis of rotation constant in time?

Consider a comoving sphere of radius $S(t) = S a(t)$ centered
on the protogalaxy.  $a(t)$ is the cosmological scale factor.
$S$ is taken to be of order but smaller than half the distance
to the nearest protogalaxy of comparable size, say one third of
that distance.  The total torque applied to the volume $V$ of
the sphere is
\begin{equation} \vec{\tau}(t) = \int_{V(t)} d^3r
~\delta\rho(\vec{r}, t)~\vec{r}\times (-\vec{\nabla} \phi(\vec{r}, t)) 
\label{torq}
\end{equation}
where $\delta\rho(\vec{r}, t) = \rho(\vec{r}, t) - \rho_0(t)$ is
the density perturbation.  $\rho_0(t)$ is the unperturbed density.
In the linear regime of evolution of density perturbations, the
gravitational potential does not depend on time when expressed in
terms of comoving coordinates, i.e.
$\phi(\vec{r} = a(t) \vec{x}, t) = \phi(\vec{x})$.  Moreover
$\delta(\vec{r}, t) \equiv {\delta \rho(\vec{r}, t) \over \rho_0(t)}$
has the form $\delta(\vec{r} = a(t) \vec{x}, t) = a(t) \delta(\vec{x})$.  
Hence
\begin{equation} \vec{\tau}(t) = \rho_0(t) a(t)^4 \int_V d^3x
~\delta(\vec{x})~\vec{x} \times (- \vec{\nabla}_x \phi(\vec{x}))~~~\ .   
\label{tt}
\end{equation}
Eq.~(\ref{tt}) shows that the direction of the torque is time independent.
Hence the rotation axis is time independent, as in the caustic ring model.
Furthermore, since $\rho_0(t) \propto a(t)^{-3}$,
$\tau(t) \propto a(t) \propto t^{2 \over 3}$
and hence $\ell(t) \propto L(t) \propto t^{5 \over 3}$.  Since    
$R(t) \propto t^{{2 \over 3} + {2 \over 9 \epsilon}}$, tidal
torque theory predicts the time dependence of Eq.~(\ref{td})
provided $\epsilon = 0.33$. This value of $\epsilon$ is in
the range, $0.25 < \epsilon < 0.35$, predicted by the evolved
spectrum of density perturbatuions and supported by the evidence
for caustic rings.  So the time dependence of the angular momentum
distribution on the turnaround sphere is also consistent with the
caustic ring model.

\section{Conclusion}

If the dark matter is axions, the phase space structure of galactic halos 
predicted by tidal torque theory is precisely, and in all respects, that of 
the caustic ring model proposed earlier on the basis of observations.  The other 
dark matter candidates predict a different phase space structure for galactic
halos.  Although the QCD axion is best motivated, a broader class of axion-like
particles behaves in the manner described here.


\begin{thebibliography}{99}

\bibitem{PDM}
For a recent review, see {\it Particle Dark Matter} edited
by Gianfranco Bertone, Cambridge University Press 2010. 

\bibitem{CABEC}
P. Sikivie and Q. Yang, Phys. Rev. Lett. 103 (2009) 111301.

\bibitem{case}
P. Sikivie, arXiv:1003:2426, to appear in Physics Letters B.

\bibitem{Baer}
H. Baer, S. Kraml, S. Sekmen and H. Summy, JHEP 0803 (2008) 056; 
H. Baer and H. Summy, Phys. Lett. B666 (2008) 5;
H. Baer, M. Haider, S. Kraml, S. Sekmen and H. Summy, JCAP 0902 (2009) 002;
H. Baer and A.D. Box, EPJC 68 (2010) 523.

\bibitem{axion}
R. D. Peccei and H. Quinn, Phys. Rev. Lett. 38 (1977) 1440 and Phys.
Rev. D16 (1977) 1791; S. Weinberg, Phys. Rev. Lett. 40 (1978) 223; 
F. Wilczek, Phys. Rev. Lett. 40 (1978) 279.

\bibitem{invis}
J. Kim, Phys. Rev. Lett. 43 (1979) 103; M. A. Shifman, 
A. I. Vainshtein and V. I. Zakharov, Nucl. Phys. B166 (1980) 493;
A. P. Zhitnitskii, Sov. J. Nucl. 31 (1980) 260;  M. Dine,
W. Fischler and M. Srednicki, Phys. Lett. B104 (1981) 199.

\bibitem{axrev}
J.E. Kim, Phys. Rep. 150 (1987) 1;
M.S. Turner, Phys. Rep. 197 (1990) 67;
G.G. Raffelt, Phys. Rep. 198 (1990) 1.

\bibitem{axdm}
J. Preskill, M. Wise and F. Wilczek, Phys. Lett. B120 (1983) 127;
L. Abbott and P. Sikivie, Phys. Lett. B120 (1983) 133;
M. Dine and W. Fischler, Phys. Lett. B120 (1983) 137.

\bibitem{axcos}
P. Sikivie, Lect. Notes Phys. 741 (2008) 19.

\bibitem{ST}   
D.V. Semikoz and I.I. Tkachev, Phys. Rev. Lett. 74 (1995) 3093 and
Phys. Rev. D55 (1997) 489.  See also: S. Khlebnikov, Phys. Rev. A66
(2002) 063606 and references therein.

\bibitem{four}
O. Erken, H. Tam, P. Sikivie and Q. Yang, to appear.

\bibitem{crdm}
P. Sikivie, Phys. Lett. B432 (1998) 139; Phys. Rev. D60 (1999) 063501.

\bibitem{inner}
A. Natarajan and P. Sikivie, Phys. Rev. D73 (2006) 023510.

\bibitem{FG}
J.A. Fillmore and P. Goldreich, Ap. J. 281 (1984) 1.

\bibitem{B}
E. Bertschinger, Ap. J. Suppl. 58 (1985) 39.

\bibitem{STW} 
P. Sikivie, I. Tkachev and Y. Wang, Phys. Rev. Lett. 75 (1995) 2911;
Phys. Rev. D56 (1997) 1863.

\bibitem{MWh}
L.D. Duffy and P. Sikivie, Phys. Rev. D78 (2008) 063508.

\bibitem{Dor}
A.G. Doroshkevitch, Astrophysics 6 (1970) 320;
P.J.E. Peebles, Ap. J. 277 (1984) 470;
Y. Hoffman and J. Shaham, Ap. J. 297 (1985) 16.

\bibitem{Kinn}
W. Kinney and P. Sikivie, Phys. Rev. D61 (2000) 087305.

\bibitem{milky}
P. Sikivie, Phys. Lett. B567 (2003) 1.

\bibitem{Mon}
A. Natarajan and P. Sikivie, Phys. Rev. D76 (2007) 023505.

\bibitem{clu}
V. Onemli and P. Sikivie, Phys. Lett. B675 (2009) 279.

\bibitem{Chemin}
L. Chemin, C. Carignan and T. Foster, Ap. J. 705 (2009) 1395.
I thank Alexey Boyarsky and Oleg Ruchayskiy for pointing 
out this paper to me.

\bibitem{num}
J.F. Navarro, C.S. Frenk and S.D.M. White, Ap. J. 462 (1996) 563;
B. Moore et al., Ap. J. Lett. 499 (1998) L5;
M. Vogelsberger, S.D.M. White, R. Mohayaee and V. Springel, MNRAS 400 (2009) 2174.

\bibitem{ttt}
G. Stromberg, Ap. J. 79 (1934) 460;
F. Hoyle, in {\it Problems of Cosmical Aerodynamics}, ed. by
J.M. Burgers and H.C. van de Hulst, 1949, p195.  Dayton, Ohio:   
Central Air Documents Office.

\bibitem{Peeb}
P.J.E. Peebles, Ap. J. 155 (1969) 2, and Astron. Ap. 11 (1971) 377.

\bibitem{Efst}
G. Efstathiou and B.J.T. Jones, MNRAS 186 (1979) 133;  
J. Barnes and G. Efstathiou, Ap. J. 319 (1987) 575;
B. Cervantes-Sodi et al., Rev. Mex. AA. 34 (2008) 87.

\bibitem{denp}
S. Weinberg, {\it Gravitation and Cosmology}, Wiley 1973;\\
S. Dodelson, {\it Modern Cosmology}, Academic Press 2003. 

\bibitem{WK}
L.M. Widrow and N. Kaiser, Ap. J. 416 (1993) L71.

\bibitem{PJ}
T. Rindler-Daller and P. Shapiro, arXiv:0912.2897.

\end{thebibliography}
\end{document}